# Testing the Quasi-temporal Gauge on the Lattice

Livio Conti[1], Claudio Parrinello[2],
Silvano Petrarca[3] and Anastassios Vladikas[4]

[1,3] Dipartimento di Fisica, Università di Roma *La Sapienza*,
Piazzale A. Moro 2, I-00185 Rome, Italy.
and
INFN, Sezione di Roma,
Piazzale A. Moro 2, I-00185 Rome, Italy.

[2] D.A.M.T.P., University of Liverpool, Liverpool L69 3BX, U.K.
(UKQCD Collaboration)

[4] INFN, Sezione di Roma 2,
Dipartimento di Fisica, Università di Roma *Tor Vergata*,
Via della Ricerca Scientifica 1, I-00133 Rome, Italy.

**Abstract**

We investigate the viability of the quasi-temporal gauge on the lattice. This is a complete gauge fixing condition that can be implemented on the lattice at a very low computational cost. As a test case, using the Clover action, we have evaluated the (gauge invariant) renormalisation constant of the non-singlet axial current, using Ward identities extracted from quark states. Our result is in reasonable but not complete agreement with previous values obtained from Ward identities both on hadronic states and on quark states in the Landau gauge. We observe large fluctuations due to lattice Gribov copies. The influence of finite volume effects is expected to be non-negligible in the case we are considering.

[1] e-mail: contil@roma1.infn.it
[2] e-mail: claudio@liverpool.ac.uk
[3] e-mail: petrarca@roma1.infn.it
[4] e-mail: vladikas@roma2.infn.it

# 1 Introduction

Gauge fixing, although in principle not necessary for the calculation of gauge invariant quantities on the lattice, is used in practice in several cases. Moreover, it is essential to the computation of gauge dependent quantities.

Typically Landau and Coulomb gauges are used. These are implemented through the iterative minimization of a suitable functional. In this way the gauge condition can be satisfied to the required precision and in many cases high accuracy is unnecessary. For example, given the gauge condition $||\partial_k A_k||^2 = 0$, where $k = 1, \ldots, 3$ or $k = 1, \ldots, 4$ for Coulomb and Landau gauge respectively, a precision of about $10^{-4} - 10^{-5}$ is adequate when computing $f_B$ in the static approximation with smeared correlation functions in the Coulomb gauge [1]. On the contrary, high precision (at least $10^{-10}$) is required for Landau gauge calculations of gluon and quark propagators [2], for identifying lattice Gribov copies [3] or for measuring the 3-gluon vertex function, in order to extract the running QCD coupling [4]. In these cases gauge fixing becomes a time consuming part of the computation, comparable to the calculation of quark propagators.

The lattice quasi-temporal gauge (QT gauge), first proposed in ref.[5] and subsequently formulated rigorously in [6], could be a low-cost alternative [1]. It is defined by fixing the Coulomb condition at a given arbitrary time $t = t_0$ and the temporal gauge at all points. It is a complete gauge fixing which, compared to the Coulomb gauge, is about $T$ times cheaper to implement on the lattice ($T$ is the length of the lattice in the time direction). A possible drawback of this gauge condition is that it is not invariant under time translations nor is free from the Gribov ambiguity. Of course the last problem is not peculiar to this gauge: Gribov copies occur in Coulomb and Landau gauges as well.

The aim of the present paper is to investigate whether the QT gauge can be used reliably in actual lattice computations. For this purpose we have calculated in this gauge the renormalisation constant of the non-singlet axial current $Z_A$, in the spirit of [8], by using Ward Identities (WI's) on quark states. $Z_A$ is suitable for our study because it is a gauge invariant quantity that has already been computed in many different ways:

1. from 1-loop lattice perturbation theory [9, 10];

2. from WI's on hadron states in a gauge invariant way [11, 12];

3. from WI's on quark states in the Landau gauge [12, 13];

4. with a non-perturbative method based on amputated quark Green functions in the Landau gauge [14].

For the third of the aforementioned determinations, a detailed study of the influence of Landau gauge Gribov copies on the measurement of $Z_A$ was performed in [13]. As our method for computing $Z_A$ coincides with the one used in [12, 13], except for changing the gauge condition from Landau to QT, we have decided to use the same ensemble of configurations that was used in these works, in order to allow a detailed comparison of the results. Also, we have analysed the rôle of the lattice Gribov copies in the QT gauge.

---

[1] Another interesting gauge, proposed in [7], can in principle also be considered.

## 2  The Temporal and Quasi-temporal Gauges

In this section we discuss the general properties of the QT gauge. First we look into the temporal gauge [15], defined by the condition

$$A_0(x) = 0 \qquad \forall x, \tag{1}$$

where $A_0(x)$ is the gauge field. In the naive path integral formulation Gauss's law is lost; eq.(1) is an incomplete gauge fixing, as it still allows time independent gauge transformations. In other words, a gauge transformation $\Omega(x)$ such that the gauge transformed field $A_0^\Omega(x)$ satisfies eq.(1) is defined up to an arbitrary gauge transformation $\Omega(\vec{x}, t_0)$ at an initial time $t_0$:

$$\Omega(x) = \mathcal{P} e^{ig \int_{t_0}^t d\tau A_0(\vec{x}, \tau)} = \Omega(\vec{x}, t_0) P(\vec{x}; t_0, t), \tag{2}$$

where $\mathcal{P}$ stands for path-ordering and $P(\vec{x}; t_0, t) = \prod_{\tau=t_0}^{t-1} U_0(\vec{x}, \tau)$ is the *Polyakov line* in the time direction from $(\vec{x}, t_0)$ to $(\vec{x}, t)$. Correspondingly, the tree level naive gluon propagator in the continuum

$$D_{lm}(x-y) = \int \frac{d^4q}{(2\pi)^4} \frac{1}{q^2} (\delta_{lm} + \frac{q_l q_m}{q_0^2}) \, e^{iq(x-y)} \tag{3}$$

suffers from a non-physical singularity $1/q_0^2$. Its regularization with a principal value prescription does not lead, to $O(g^4)$ in perturbation theory, to the correct exponentiation of the Wilson loop [16]. One way of solving all these problems at once is to integrate on time independent gauge transformations [17]. This operation is equivalent to restricting the space of states to those satisfying Gauss's law which is then imposed weakly on the physical states.

As an alternative procedure one can enforce the Coulomb condition at a given time $t = t_0$ [5, 6]:

$$\vec{\partial} \cdot \vec{A}(\vec{x}, t_0) = 0 \quad \forall(\vec{x}, t_0). \tag{4}$$

Eqs.(1) and (4) define the QT gauge. This is a complete gauge fixing, Gauss's law is satisfied at $t = t_0$ and hence at all times and the problematic $q_0^2$ pole is removed from the tree level gluon propagator, which has the form [6] [2]:

$$< T(A_l^a(\vec{x}, x_0) A_m^b(\vec{y}, y_0)) > = -\frac{\delta^{ab}}{2} \int \frac{d\vec{q}}{(2\pi)^3} e^{i\vec{q}(\vec{x}-\vec{y})} \left[ (\delta_{lm} - \frac{q_l q_m}{|\vec{q}|^2}) \frac{e^{-|\vec{q}||x_0-y_0|}}{|\vec{q}|} + \right.$$
$$\left. -\frac{q_l q_m}{|\vec{q}|^2}(|x_0 - y_0| - |x_0 - t_0| - |y_0 - t_0|) \right]. \tag{5}$$

Note that, because of the Coulomb condition at $t_0$, the QT gauge breaks translational invariance in the time direction. This is why gauge dependent quantities, like the propagator of eq.(5), are not invariant under time translations. (Translational invariance holds for (5) when the time ordering is either $x_0 \leq t_0 \leq y_0$ or $y_0 \leq t_0 \leq x_0$.)

Analytic calculations with the above propagator are hard to perform. However, in [6] the correct exponentiation of the Wilson loop to $O(g^4)$ in the QT gauge was explicitly demonstrated.

---
[2] For a different calculation of this quantity see also [18].

## 3 The Quasi-temporal Gauge on the Lattice

We now turn to the implementation of the QT gauge on the lattice. Since at the time $t_0$ one needs to set $\partial_k A_k = 0$, $k = 1, \ldots, 3$, we first review briefly the implementation of the Coulomb gauge on the lattice.

Given a thermalized Monte Carlo configuration $U$ on an $L^3 \times T$ lattice with periodic boundary conditions, on each timeslice $t$ one can define the functional [19]

$$F[U^\Omega](t) \equiv -\frac{1}{L^3} Re\ Tr \sum_{k=1}^{3} \sum_{n=1}^{L^3} U_k^\Omega(n, t), \qquad (6)$$

where $U_k^\Omega(n,t) \equiv \Omega(n,t) U_k(n,t) \Omega^\dagger(n+\hat{k},t)$ represents a gauge transformed $SU(3)$ link, $\Omega(n,t)$ being a gauge transformation matrix. A link configuration $U^\Omega$ which satisfies the lattice analogue of the Coulomb gauge condition at the time $t$ can be obtained by finding a minimum of $F(t)$ with respect to gauge transformations $\Omega$. More generally, it can be shown that all the extrema of $F(t)$ with respect to $\Omega$ correspond to configurations which satisfy the Coulomb gauge condition $\partial_k A_k^\Omega = 0$ in discretized form. The existence of more than one minimum is related to the presence of Gribov copies.

It is interesting to note that on the lattice one has in general more classes of minima than in the continuum [20, 21]. The issue of distinguishing between continuum - like minima and lattice artifacts is certainly an interesting challenge, but we shall ignore it in the following. We adopt the pragmatic point of view that in a numerical simulation both kinds of minima turn up.

On the lattice, the minima of $F(t)$ can be found numerically by iteration. The typical minimization algorithm sweeps through all lattice sites $n$ at a fixed time $t$ and performs local gauge transformations which minimize $F(t)$ with respect to the local gauge group.

Obviously, in order to implement the Coulomb gauge on the entire lattice one should repeat the above procedure for all the timeslices, i.e. $T$ times. On the other hand, in the QT gauge, we only need to impose the Coulomb gauge on a single timeslice $t_0$. In this work we have chosen $t_0 = 0$, i.e. the first timeslice of the lattice. In our notation the time index $t$ varies from 0 to $(T-1)$.

Once the Coulomb condition holds at $t = t_0$, the temporal gauge (1) can be trivially imposed by visiting sequentially each timeslice and gauge transforming the temporal links $U_0(n,t)$ into the unit group element. On a periodic lattice, this can be done for all but one time $t_f$, so that the temporal links $U_0(n, t_f)$, rather than being unity, end up carrying the value of the Polyakov loops.

In this work we have chosen $t_f = T/2 - 1$. We have completed the QT gauge by fixing first the temporal gauge on the timeslices $t = 1, \ldots, (T/2 - 1)$, sweeping the lattice in the forward direction, which yields $U_0(n, t-1) = 1$ for $t = 1, \ldots, (T/2-1)$. Then the gauge is fixed on the timeslices $t = (T-1), \ldots, T/2$, sweeping the lattice backwards, giving $U_0(n,t) = 1$ for $t = (T-1), \ldots, T/2$. Thus, the Polyakov line is now between the timeslices $(T/2 - 1)$ and $T/2$.

In this way, once the QT gauge is fixed, we end up with a lattice which is symmetric in the time direction around $t = 0$. The discontinuity due to the Polyakov line is placed at the greatest possible distance from $t = 0$ where the signal is killed anyway by statistical fluctuations due to the strong damping of Green functions.

Since the computational cost of fixing the temporal gauge is negligible, it follows that the QT gauge is roughly $T$ times faster to implement than the Coulomb gauge. This may be a considerable saving in cases where high accuracy is required.

The QT gauge suffers, like other gauges, from the Gribov ambiguity. Gribov copies can be generated when fixing the Coulomb gauge at $t = t_0$. When the temporal gauge is subsequently implemented, the transformation $\Omega$, necessary to perform the gauge rotation, will depend for all values of $t$ on possible Gribov copies, so that their existence affects the gauge fixing at all timeslices. This is to be contrasted to the situation in the Coulomb gauge, where gauge fixing is an independent process on each timeslice and Gribov copies on one timeslice do not depend on what happens at other times [3].

## 4  The Measurement of $Z_A$

In order to test the feasibility of lattice non-perturbative calculations in the QT gauge, we have calculated the renormalisation constant $Z_A$ of the axial current by using WI's on quark states. As we have pointed out in the introduction, this quantity has already been measured with several different methods, using both the Wilson and the Clover actions. We adopted the latter action [22, 23] which is free from $O(a)$ discretization errors. We only give an outline of the method and we refer the reader to [12] for the details.

The gauge fixed calculation of $Z_A$ relies on the following WI for quark Green functions

$$2\rho Tr\left[\int d^4x \int d\vec{y} < u_\alpha(y)\,(\bar{u}(x)\gamma_5 d(x))\,\bar{d}_\beta(0) > \right] =$$
$$= \left(\frac{1}{Z_A} - \rho a\right) Tr\left[\int d\vec{y} < ((\gamma_5 d(y))_\alpha \bar{d}_\beta(0) + u(y)_\alpha(\bar{u}(0)\gamma_5)_\beta) > \right]. \quad (7)$$

In eq.(7) we work explicitly with up ($u$) and down ($d$) quark fields with spinor indices $\alpha$ and $\beta$. The trace is over colour indices. The value of $\rho$ is obtained from the following ratio of 2-point correlation functions:

$$2\rho = \frac{\partial_4 \int d\vec{y} < \mathcal{A}_4(\vec{y},t_y)\mathcal{P}_5^\dagger(\vec{0},0) >}{\int d\vec{y} < \mathcal{P}_5(\vec{y},t_y)\mathcal{P}_5^\dagger(\vec{0},0) >} = \frac{2m}{Z_A}, \quad (8)$$

where

$$\begin{aligned}\mathcal{A}_\mu(x) &\equiv \bar{u}(x)\gamma_\mu\gamma_5 d(x) \\ \mathcal{P}_5(x) &\equiv \bar{d}(x)\gamma_5 u(x)\end{aligned} \quad (9)$$

are the axial current and the pseudoscalar density and $m$ is the bare quark mass. As can be seen from eq.(8), $\rho$ is gauge invariant. Note that the $\rho a$ term on the r.h.s. of eq.(7) is peculiar to the Clover action and arises from the field redefinitions necessary in this formulation [23]. In this work we use the "improved - improved" propagators of [10, 24].

In order to compute $Z_A$ we evaluate (8) and the traces in (7) as functions of $t_y$. In our case, the (gauge dependent) traces have been evaluated in the QT gauge. One can then solve eq.(7) for $Z_A$ as a function of $t_y$. In the lattice simulations a plateau in $t_y$ is typically seen, from which the estimate for $Z_A$ is obtained.

Actually, barring contact terms, the WI of eq.(7) is expected to hold at all times. In practice, as it was already noted in [12], the presence of such terms affects the behaviour of the curve near $t = 0$.

## 5  Numerical Results

We have used data obtained from a Monte Carlo simulation on a $16^3 \times 32$ lattice at $\beta = 6.0$, with the Clover action of $SU(3)$ gauge theory, in the quenched approximation. An 8-hit Metropolis algorithm was used to generate an ensemble of 18 configurations, each separated by 1000 sweeps, after an initial thermalization of 3000 sweeps. The accuracy of the Coulomb gauge fixing at $t = 0$ is determined by the requirement that $F$ be minimized within a precision of $\Delta F/F < 10^{-10}$, where $\Delta F$ is the change in $F$ between two successive gauge fixing sweeps. To increase the convergence of the algorithm we used the overrelaxation method [25], setting the overrelaxation parameter to $\omega = 1.72$. The whole calculation (and, in particular, the gauge fixing) was done in double precision (64 bit). Typically, about 100 gauge fixing iterations were needed in order to reach the required precision. In our no-tation ($T = 32$, $0 \leq t \leq 31$ and $t_f = 15$) the Polyakov line is between $t = 15$ and $t = 16$.

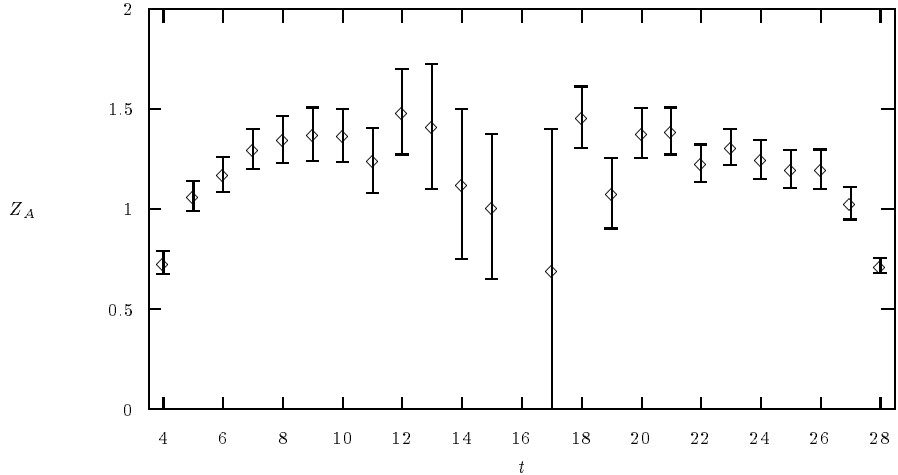

Figure 1: $Z_A$ as a function of time in the QT gauge. The value at $t = 16$ is not shown, as it is affected by large fluctuations.

In this exploratory study we have used quark propagators at a single value of the hopping parameter, $\kappa = 0.1425$, which corresponds to a pion mass of about 900 MeV. The value of $\rho a$ is about 0.05. For light quarks, the dependence of the measured renormalisation constants on the quark mass is very mild [24, 26].

In order to enhance the signal, we have averaged eq.(7) over four contributions, corresponding to the values of the Dirac indices $(\alpha, \beta) = (1,3), (3,1), (2,4)$ and $(4,2)$. These were found to yield the clearest signal (the same is true in the Landau gauge [12]). The statistical errors have been obtained with the jacknife method by decimating one configuration at a time.

Our estimate for $Z_A$ is shown in Fig.1 as a function of time $t$. Except for the behaviour near the origin, attributed to the presence of contact terms, in the intervals $7 \leq t \leq 11$ and $22 \leq t \leq 26$, plateaux appear to settle in. The quality of the signal worsens at $t \sim T/2$ due to the exponential decay of the correlation functions. At times $t = 15$ and $16$, where the Polyakov line is located, the signal is completely lost. We estimate

$$Z_A = 1.33 \pm 0.13 \qquad Z_A = 1.24 \pm 0.10 \qquad (10)$$

from the plateaux at the left and at the right of $t \sim T/2$ respectively. As our choice of imposing the Coulomb condition at $t_0 = 0$ and the way we fix the temporal gauge both preserve the symmetry of the quark propagators $S(x,0)$ around $t=0$, we can average the above results to obtain

$$Z_A = 1.28 \pm 0.11 . \qquad (11)$$

Since $Z_A$ is gauge independent, it is of particular significance to compare this result with the Landau gauge one, as they have both been obtained from the same WI. In [12] $Z_A$ was calculated at the same $\beta$ and $\kappa$ values, on the same ensemble of 18 configurations. The calculation was performed both with the gauge invariant procedure based of WI's on hadron states, and with the gauge dependent procedure used in this work, but employing the Landau gauge. The comparison of the values obtained for $Z_A$ as a function of time in the Landau and in the QT gauge is shown in Fig.2 . Although the behaviour of $Z_A$ in both gauges is qualitatively similar, the central values of the QT results are shifted upward with respect to the Landau ones. The QT results also display larger statistical fluctuations and a shorter plateau. This is reflected in the comparison of the result of eq.(11) to those quoted in [12]: $Z_A = 1.09 \pm 0.03$ (gauge invariant method) and $Z_A = 1.14 \pm 0.08$ (Landau gauge). We consider the gauge invariant value as the best estimate of $Z_A$ on this ensemble; it is free from the problems associated with gauge fixing and is obtained from a much wider and more stable plateau. We note that the QT estimate, although compatible with the Landau one, compares rather poorly to the gauge invariant one.

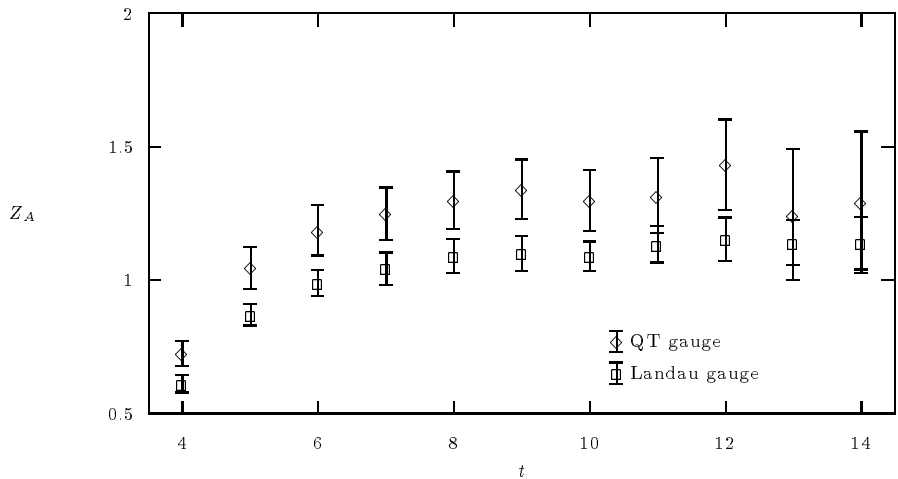

Figure 2: Comparison of $Z_A$ calculated in the QT gauge and in the Landau gauge [12]. The Landau results were obtained by symmetrising the correlation functions with respect to the origin, whereas the QT ones are obtained by averaging the value of $Z_A$ around the origin.

For completeness, we also quote the results of [13], obtained on a enlarged ensemble of 36 configurations at the same lattice volume, $\beta$ and $\kappa$. From the gauge independent WI's they obtain $Z_A = 1.06 \pm 0.02$, whereas from the WI's on quark states in the Landau gauge $Z_A = 1.08 \pm 0.05$. The latter result has been obtained by averaging over 6 Gribov copies for each thermalized configuration. This procedure has been introduced in a different framework in ref.[27]. Finally, we also quote from [10] the result $Z_A = 0.97$, obtained in boosted lattice perturbation theory in the spirit of ref.[28].

The discrepancy between the determination of $Z_A$ in the QT gauge and the other estimates reported above could be attributed to the breaking of translational invariance in time direction inherent in this gauge. It is conceivable that the loss of this symmetry may cause appreciable finite volume effects causing large systematic errors in the measurement.

Another source of error may be the presence of lattice Gribov copies, which, as pointed out in [13], also cause an increase of the statistical error. According to the discussion in Sect.(3), Gribov copies in the QT gauge affect the gauge transformations $\Omega$ over the whole lattice and may thus have a significant effect on the determination of $Z_A$.

We have studied the effect of the existence of lattice Gribov copies on $Z_A$ following ref.[13]. For each thermalized configuration three lattice Gribov copies have been generated by performing random gauge transformations before fixing the QT gauge. We have obtained in this way three measurements of $Z_A$, using a different Gribov copy for each configuration. Besides the result given in eq.(11), we obtain for the three copies:

$$Z_A = 1.23 \pm 0.13$$
$$Z_A = 1.30 \pm 0.13$$
$$Z_A = 1.23 \pm 0.16 \,. \qquad (12)$$

Although the results of eqs.(11) and (12) are compatible, there are big fluctuations between results obtained on different copies. Like in [13] our best estimate for $Z_A$ is obtained by averaging over the four values:

$$Z_A = 1.26 \pm 0.13 \,. \qquad (13)$$

This analysis shows that the presence of lattice Gribov copies in the QT gauge is responsible of visible fluctuations on the $Z_A$ value, at the same level of what is observed in the Landau case, [13]. Nevertheless the larger error found in the QT gauge, compared to the Landau case, cannot be attributed completely to this effect.

## 6 Conclusions

We have investigated the feasibility of lattice computations in the QT gauge. The cost of gauge fixing is reduced by approximately a factor of $T$ with respect to conventional gauge fixings (Coulomb or Landau). The renormalisation constant of the axial current $Z_A$ has been measured. Our final numerical result agrees only roughly with other estimates, within errors. A detailed comparison of our results to what is obtained from WI's on hadron states or quark states in the Landau gauge shows, in fact, that our numbers are systematically characterised by a higher central value and a larger statistical error. We have shown that these features can be partly related to the lattice Gribov ambiguity. Also, the breaking of translational invariance in the time direction may be responsible for an enhancement of systematic errors from finite volume effects. We are currently investigating such effects by repeating the calculation at a larger volume.

We are extremely grateful to G.C.Rossi and M.Testa for an early participation to this work and for many illuminating discussions; many useful discussions with J.P.Leroy, G.Martinelli and B.Taglienti are also gratefully acknowledged. C.P. acknowledges support from PPARC through an Advanced Fellowship.